\documentclass{article}
\PassOptionsToPackage{square, numbers}{natbib}
\usepackage[final]{neurips_2024}

\usepackage[utf8]{inputenc}
\usepackage[T1]{fontenc}
\usepackage{amsmath}
\usepackage{amsfonts}
\usepackage{amssymb}
\usepackage{graphicx}
\usepackage{hyperref}
\usepackage{booktabs}
\usepackage{array}
\usepackage{multirow}
\usepackage{float}
\usepackage{subcaption}
\usepackage{url}
\usepackage{xcolor}
\usepackage{geometry}
\usepackage{enumitem}

\definecolor{primaryblue}{RGB}{0, 114, 178}
\definecolor{secondarygreen}{RGB}{0, 158, 115}
\definecolor{accentred}{RGB}{213, 94, 0}

\hypersetup{
    colorlinks=true,
    linkcolor=primaryblue,
    citecolor=secondarygreen,
    urlcolor=accentred
}

\title{Evaluation and Benchmarking Suite for Financial Large Language Models and Agents}

\author{
  Shengyuan Lin$^{1,2}$, Kaiwen He$^{1,3}$, Jaisal Patel$^3$, Qinchuan Zhang$^3$, Chris Ding$^3$, James Tang$^4$ \\
  \textbf{Keyi Wang$^1$, Yupeng Cao$^5$, Yan Wang$^6$, Kairong Xiao$^7$, Vincent Caldeira$^8$} \\
  \textbf{Matt White$^{9,10}$, Xiao-Yang Liu Yanglet$^1$\thanks{Corresponding author.}} \\
  $^1$SecureFinAI Lab, Columbia University,
  $^2$Carnegie Mellon University, 
  $^4$Boston College,\\
  $^3$Rensselaer Center of Open Source, Rensselaer Polytechnic Institute, 
  \\$^5$Stevens Institute of Technology, $^6$The FinAI, $^7$Business School, Columbia University, \\
  $^8$Red Hat, $^9$PyTorch Foundation, $^{10}$Linux Foundation \\
  \texttt{shengyu3@andrew.cmu.edu},~   
  \texttt{XL2427@columbia.edu}
}

\begin{document}

\maketitle

\begin{abstract}
Over the past three years, the financial services industry has witnessed Large Language Models (LLMs) and agents transitioning from the exploration stage to readiness and governance stages. Financial large language models (FinLLMs), such as open FinGPT \cite{liu2023fingpt} and proprietary BloombergGPT \cite{wu2023bloomberggpt}, have great potential in financial applications, including retrieving real-time data, tutoring, analyzing sentiment of social media, analyzing SEC filings, and agentic trading. However, general-purpose LLMs and agents lack financial expertise and often struggle to handle complex financial reasoning. This paper presents an evaluation and benchmarking suite that covers the lifecycle of FinLLMs and FinAgents. This suite led by SecureFinAI Lab includes an evaluation pipeline and a governance framework collaborating with Linux Foundation and PyTorch Foundation, a FinLLM Leaderboard with HuggingFace, an AgentOps framework with Red Hat, and a documentation website with Rensselear Center of Open Source. Our collaborative development evolves through three stages: FinLLM Exploration (2023), FinLLM Readiness (2024), and FinAI Governance (2025). The proposed suite serves as an open platform that enables researchers and practitioners to perform both quantitative and qualitative analysis of different FinLLMs and FinAgents, fostering a more robust and reliable FinAI ecosystem.
\end{abstract}

% \begin{IEEEkeywords}
% Financial Large Language Models, Benchmarking, Evaluation Framework, Financial AI, LLM Lifecycle, Governance
% \end{IEEEkeywords}

\section{Introduction}
\label{sec:introduction}

Business and finance are high-stakes domains for AI applications. Typical financial applications include retrieving financial data, analyzing earning conference calls (ECC), and analyzing sentiment of social media. Evaluating and benchmarking the LLMs' gap in financial knowledge and reasoning is critical for building financial agents. It helps pave the way for democratizing financial intelligence to the general public and foster a more robust and reliable FinAI ecosystem.

However, this interdisciplinary field still lacks a systematic evaluation framework that covers lifecycle of FinLLMs and FinAgents. There exist works that assessed how LLMs perform on financial exams and financial benchmarks, such as FinBen \cite{xie2024finben} and FinanceBench \cite{islam2023financebench}. However, these benchmarking works are still an early stage exploration for LLMs, leaving a gap between industry readiness and governance. This gap is particularly critical because the financial domain poses unique challenges that demand evaluation methodologies grounded in deep domain expertise.

To bridge the gap, our goal is to \textbf{explore financial use cases} that reveal both the potential and the limitations of current LLMs across diverse financial scenarios. Currently, the lack of standardized evaluation methods in financial AI creates barriers in which different institutions use varying metrics and evaluation criteria, making it difficult to assess model performance objectively. To address this, we \textbf{promote a de facto standard in the financial services industry} through benchmarking and evaluation frameworks. 

\section{Overview of Evolving LLM Lifecycle in Finance}
\label{sec:overview}

% The evolution of Large Language Models (LLMs) in the financial domain has followed a distinct trajectory characterized by three major stages, each representing different levels of maturity and focus areas. This lifecycle framework provides a systematic way to understand the development, challenges, and future directions of financial AI systems.

\begin{figure*}[t]
    \centering
    \includegraphics[width=\textwidth]{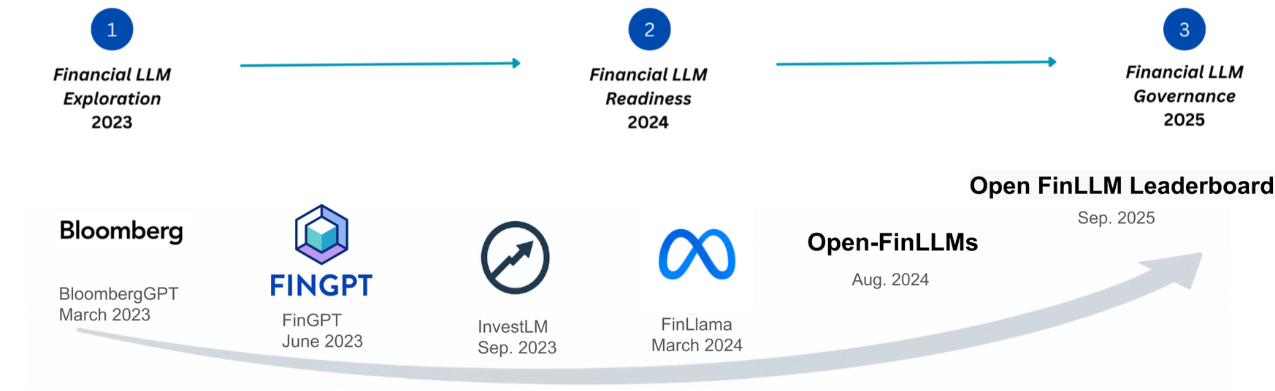}
    \caption{The development of FinLLMs and agents consists of three stages: FinLLM Exploration (2023), FinLLM Readiness (2024), and FinAI Governance (2025), along with a reference timeline. % of notable Financial Large Language Models showing the progression from foundational models like BloombergGPT to more recent and multimodal systems.
    }
    \label{fig:lifecycle_stages}
\end{figure*}

Over the past three years, SecureFinAI Lab at Columbia University has organized bi-weekly meetings with Linux Foundation and PyTorch Foundation and $20+$ FinTech companies (industry partners). The topic theme has evolved from FinLLM Exploration to FinLLM Readiness and FinAI Governance, as shown in Fig. \ref{fig:lifecycle_stages}.
\begin{itemize}[leftmargin=*]
    \item \textbf{FinLLM Exploration (2023)} studied the promising applications of financial large language models. Researchers and practitioners focused on developing foundational models such as the proprietary BloombergGPT \cite{wu2023bloomberggpt} and the open-source FinGPT \cite{liu2023fingpt}, showcasing their capabilities in financial tasks such as question answering and sentiment analysis. We created a standard evaluation pipeline and an  \href{https://huggingface.co/spaces/finosfoundation/Open-Financial-LLM-Leaderboard}{Open Financial LLM Leaderboard} held on HuggingFace.
    \item \textbf{FinLLM Readiness (2024)}, represents a shift toward systematic evaluation and benchmarking. In this stage researchers encouraged the development of comprehensive benchmarks like FinBen \cite{xie2024finben} and MultiFinBen \cite{peng2025multifinben}, as well as the emergence of financial agents and specialized APIs. The focus shifted from basic capability demonstration to performance evaluation and real-world applications.
    \item \textbf{FinAI Governance (2025)}, addresses the critical challenges of responsible AI deployment in financial contexts. This stage focuses on identifying and mitigating risks associated with financial LLMs, including hallucination, security vulnerabilities, and regulatory compliance issues. The key initiative is the \href{https://air-governance-framework.finos.org/}{AI Governance Framework}.
\end{itemize}

Please note that \textbf{benchmarking} ranks LLMs performance relative to others, while \textbf{evaluation} determines if a FinAgent meets tailored expectations.

\section{FinLLM Exploration: Benchmarking through an Open Leaderboard}
\label{sec:exploration}

In 2023, researchers began exploring how large language models could be used in finance. This was the beginning of financial LLMs. People wanted to see if these models could understand financial language and help with financial tasks.

\subsection{Timeline for FinLLMs}
\label{sec:timeline}

The evolution of Financial Large Language Models (FinLLMs) can be traced through a timeline of notable milestones. Fig.~\ref{fig:lifecycle_stages} illustrates the rapid shift from foundational financial-domain pretraining to more advanced and multimodal financial AI systems. Beginning with foundational models like BloombergGPT in early 2023, we observe a progression through various specialized financial models, including FinGPT, domain-adapted models, and evolving toward AI governance frameworks. 

\subsection{Financial Tasks}
\textbf{Financial Tasks with Multimodal Data} We compare FinLLMs across multiple task categories including information extraction (IE), textual analysis (TA), question answering (QA), text generation (TG), risk management (RM), forecasting (FO) and Decision-Making (DM). The current $42$ financial datasets are organized into seven categories, as given in Table \ref{tab:task_categories}. The educational documents of the open FinLLM leaderboard is available on this \href{https://finllm-leaderboard.readthedocs.io/en/latest/}{website}.

\begin{table*}[b]
\centering
\caption{Financial tasks in the open FinLLM leaderboard.}
\label{tab:task_categories}
\begin{tabular}{|l|l|}
\hline
\textbf{Category}            & \textbf{Financial Tasks}                                                                        \\ \hline
\textbf{Information Extraction (IE)}  & Named Entity Recognition (NER), Relation Extraction, Causal Classification.         \\ \hline
\textbf{Textual Analysis (TA)}        & Sentiment Analysis, Classification, Argument Unit Classification.               \\ \hline
\textbf{Question Answering (QA)}      & Answering financial questions from datasets like FinQA and TATQA.                              \\ \hline
\textbf{Text Generation (TG)}         & Summarization of financial texts (e.g., reports, filings).                          \\ \hline
\textbf{Risk Management (RM)}         & Credit Scoring, Fraud Detection, evaluating financial risks.                        \\ \hline
\textbf{Forecasting (FO)}             & Stock Movement Prediction based on financial news and social media.                            \\ \hline
\textbf{Decision-Making (DM)}         & Simulating decision-making tasks, e.g., M\&A transactions, trading tasks.                       \\ \hline
\end{tabular}
\end{table*}

\textbf{Towards financial readiness} Our goal would be to build an open community that pushes financial AI to be ready for real world applications and to build a gateway between academia and industry. The Open FinLLM Leaderboard is similar to established industry standards such as MCP and MOF. We set the benchmark for financial AI readiness, ensuring that innovations in financial language models are both practical and impactful. 

\begin{figure*}[t]
    \centering
    \includegraphics[width=\textwidth]{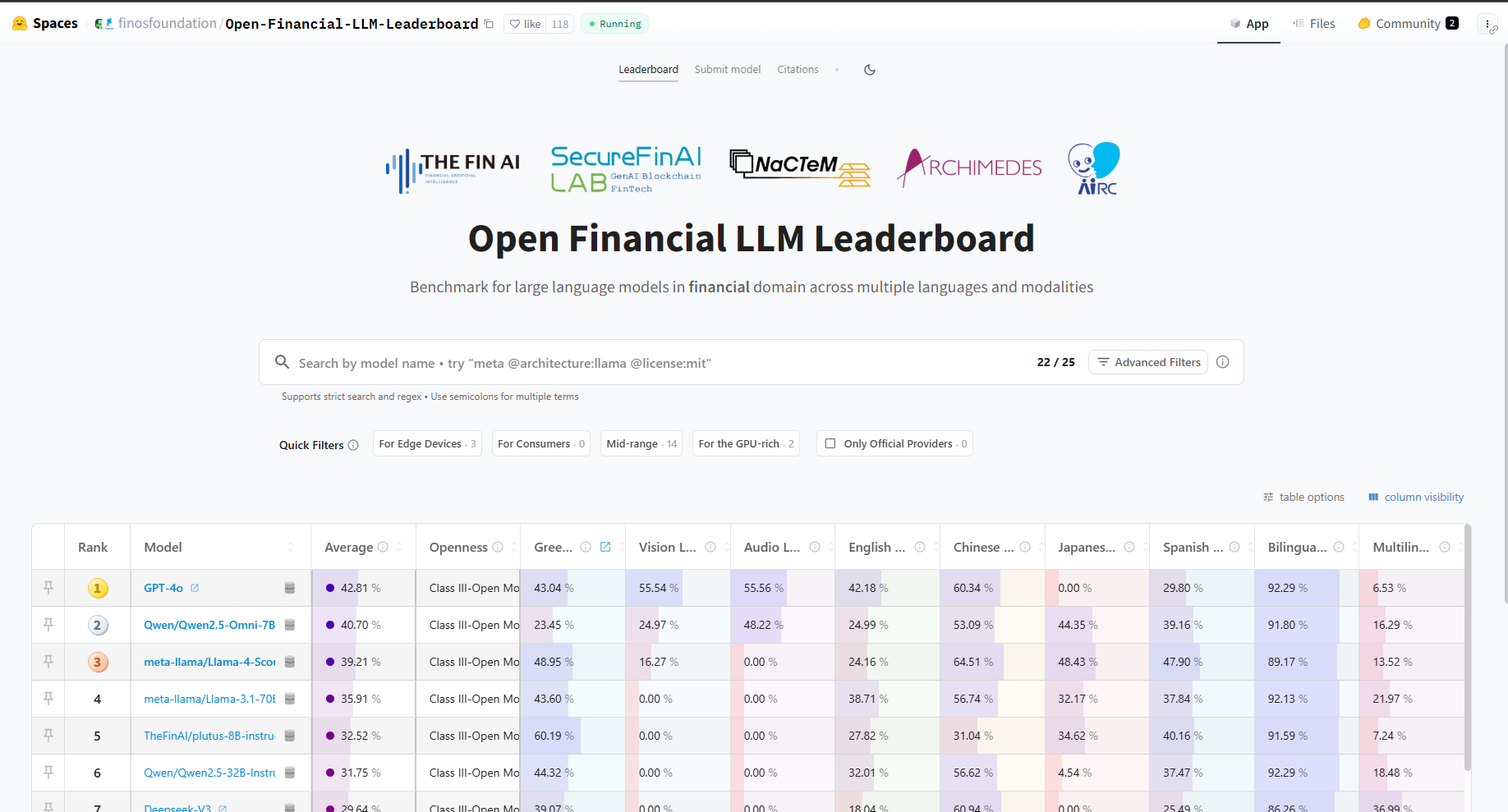}
    \caption{Interface of the Open Financial LLM Leaderboard on Huggingface, showing ranked leaderboard table with multilingual and multimodal financial task scores.}
    \label{fig:leaderboard_ui}
\end{figure*}

\subsection{Open FinLLM Leaderboard}
\label{sec:leaderboard}

\href{https://huggingface.co/spaces/finosfoundation/Open-Financial-LLM-Leaderboard}{The Open FinLLM Leaderboard} is an open platform that evaluates and compares FinLLMs and FinAgents across a wide spectrum of financial tasks. It is a collaborative project with the Linux Foundation and Hugging Face. This leaderboard provides a transparent and standardized framework that ranks models based on their (multimodal) performance in areas such as financial reporting, sentiment analysis, and stock prediction. It also serves as an open platform for the community to evaluate, interact with, and compare FinLLMs and FinAgents in real-world scenarios. Beyond numeric scores, we showcase the integration with the FinGPT Search Agent \cite{tian2024customized}, a promising use case of a personalized financial advisor. Users can explore, interact with, and compare models through demos. Additionally, we encourage contributions of models, datasets, and tasks to keep the leaderboard dynamic and responsive to the evolving needs of the financial industry. The leaderboard is continuously evolving, ensuring that it remains up-to-date with the latest FinLLMs and agents and adapts to more professional-level financial tasks. 

As financial institutions adopt third-party models, the concern with "open-washing", where models claim to be open but withhold critical training data or licensing, poses a significant compliance risk. To address this, we integrate the Model Openness Framework \cite{mof2024white} \footnote{\url{https://isitopen.ai/}}  directly into the Open FinLLM Leaderboard. By filtering the leaderboard through the various Model Openness Framework classes, we enable financial institutions to select models not just on capability (benchmark scores) but on compliance readiness.

\subsection{Leaderboard Demo}
\label{subsec:leaderboard_ui}

Fig.~\ref{fig:leaderboard_ui} shows the user interface of the 
\href{https://huggingface.co/spaces/finosfoundation/Open-Financial-LLM-Leaderboard}{Open Financial LLM Leaderboard}. 
The leaderboard on Huggingface is designed to show transparent and comprehensive evaluation results for financial LLMs. 

\textbf{Search Bar and Filtering Tools.} Users can search by model name, architecture, and license expression.  

\textbf{Leaderboard Table.} The central part is a table containing: rank, model name, average score across all tasks, openness class (Class I, II, III under MOF in Section \ref{sec:MOF}),

\section{FinLLM Readiness: Evaluating Financial Agents }
\label{sec:benchmarking}

\begin{figure}[t]
    \centering
    \includegraphics[width=\textwidth]{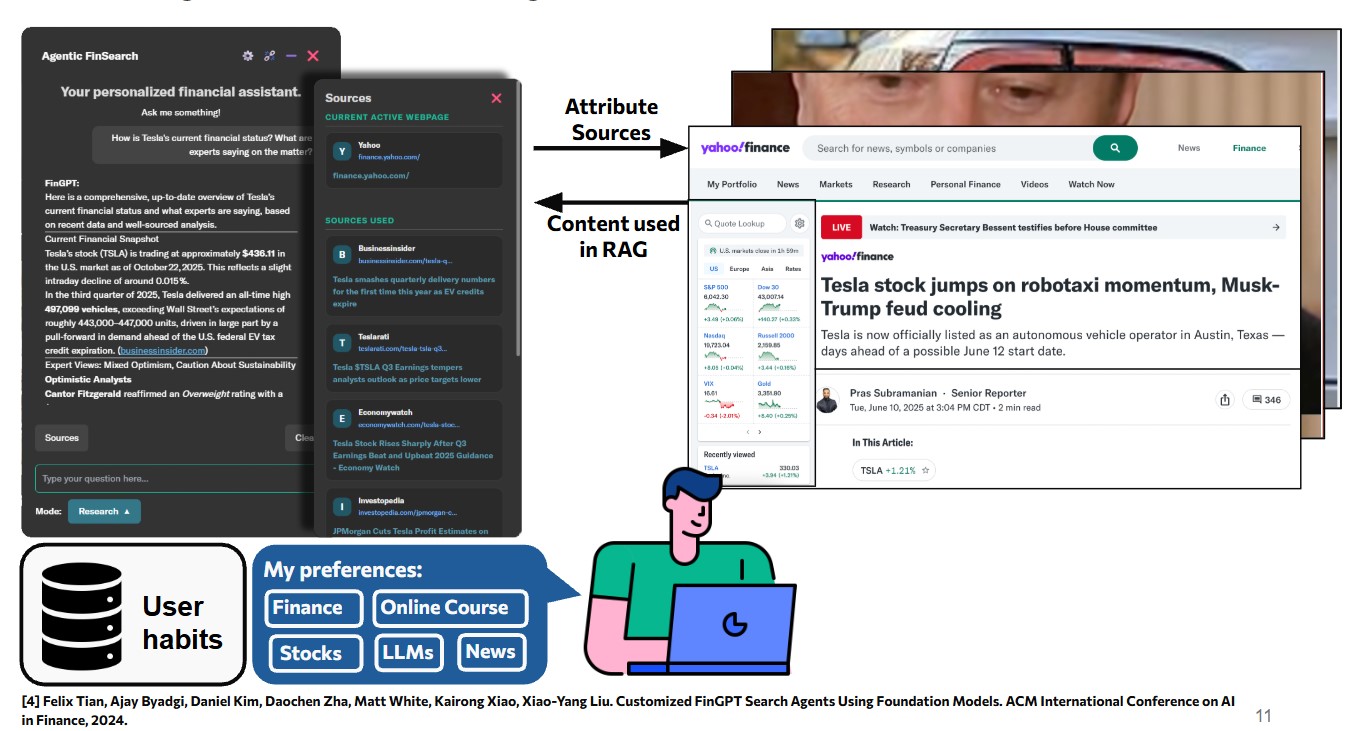}
    \caption{Agentic FinSearch for analyzing Tesla stock news from CNBC. The agent scrapes the website, extracts content, and provides key findings with source attribution.}
    \label{fig:search_agent}
\end{figure}

\subsection{Agentic FinSearch}

Our agentic FinSearch \cite{tian2024customized} can analyze financial websites like YahooFinance and Bloomberg with the integration of retrieval augmented generation (RAG). This agent autonomously navigate complex websites, perform high-quality analysis, and execute multi-step data retrieval. Unlike Blroomberg terminal that requires extensive manual operations, search an agent can autonomously perform tasks such as real-time market data collection, news sentiment analysis, and competitive intelligence gathering. In time-sensitive financial tasks like trading, agentic FinSearch with quick respond and high accuracy can directly improve profitability.

%The Agentic FinSearch is a customizable AI search agent that can scrape and analyze websites, fetch user-specified websites, search local files, and verify sources. It offers two key features: \textbf{personalization} and \textbf{intelligent search}. The personalization feature learns user habits and preferences from a dynamic database and user feedback, allowing users to maintain a list of preferred websites and select specific LLMs. The intelligent search capability enables the agent to scrape websites or search local files based on user queries, then extract relevant content through RAG and answer queries while providing source attribution.

Fig.~\ref{fig:search_agent} illustrates a use scenario where a user reading a Tesla article on CNBC can directly open the YahooFinance and ask agentic FinSearch: "based on this article, what are some key findings relates to Tesla?" The agent scrapes the current website CNBC, extracts relevant content, and use LLMs to summarize key findings of Tesla. Users can add authentic websites into the list of preferred website.

\textbf{Key Features of Agentic FinSearch}. Compared to general-purpose AI search agents, our agentic FinSearch offers three key features
\begin{itemize}[leftmargin=*]
    \item \textbf{Air-gapped Deployment}. In March 2023, a \href{https://openai.com/index/march-20-chatgpt-outage/?utm_source=chatgpt.com}{security incident} exposed some ChatGPT users' chat history, raising serious concerns about the leakage risk of users' private financial data. FinGPT can address such vulnerability through air-gapped deployment. It can ensure complete isolation from external networks and prevent data leakage by implementing a local solution for financial institutions to handle sensitive client information and trading strategies.
    \item \textbf{Higher Numerical Accuracy}. \href{https://www.searchenginejournal.com/googles-ai-fails-at-43-of-finance-queries-study-finds/530542/?utm_source=chatgpt.com}{Google AI Overview} has demonstrated significant accuracy issues in financial contexts: 43\% of finance summaries are inaccurate or misleading, and 57\% of life insurance information is incorrect. FinGPT can achieve 85\% accuracy in financial numerical reasoning tasks, which is higher than Perplexity's 55\%. Numerical accuracy is crucial for financial applications in terms of trading decisions and regulatory compliance, where a small mistake can leads to significant losses.
    \item \textbf{Lower Hallucination \& Misinformation}. General-purpose AI systems have produced costly errors in financial contexts. \href{https://www.kiplinger.com/personal-finance/insurance/google-ai-life-insurance-overview-wrong-57-percent-study?utm_source=chatgpt.com}{Google's AI Overview} provided incorrect insurance and Medicare guidance, and Google lost \$100 billion in market value at Bard's launch due to a factual error in its demonstration. FinGPT can mitigate risks by fine-tuning with high-quality financial data, retrieval-augmented generation (RAG), and fact-checking responses.
\end{itemize}

\subsection{Tutor Agent}
 The Tutor Agent is an intelligent financial education assistant powered by pretrained large language models that aims to democratize financial education by providing affordable, scalable, and high-quality learning support to the general public. These agents can assist with exam preparation for credit risk management, professional development for learning financial terms, research advising for independent study, and CFA exam preparation with 24/7 access to expert-level explanations and tailored practice questions.

The motivation behind financial AI tutors addresses three key challenges in financial education: \textbf{affordability}, \textbf{scalability}, and \textbf{democratization}. Traditional tutoring services are expensive and inaccessible to many students. AI tutors provide cost-effective deployment using pretrained LLMs, reducing the cost per student through automation. They offer scalable solutions by serving millions of users simultaneously, available 24/7 without geographical limitations, while maintaining consistent quality across all users. Most importantly, they democratize education by breaking down barriers to financial knowledge and enabling self-learning for people from all backgrounds.

\begin{figure}[htbp]
    \centering
    \includegraphics[width=\textwidth]{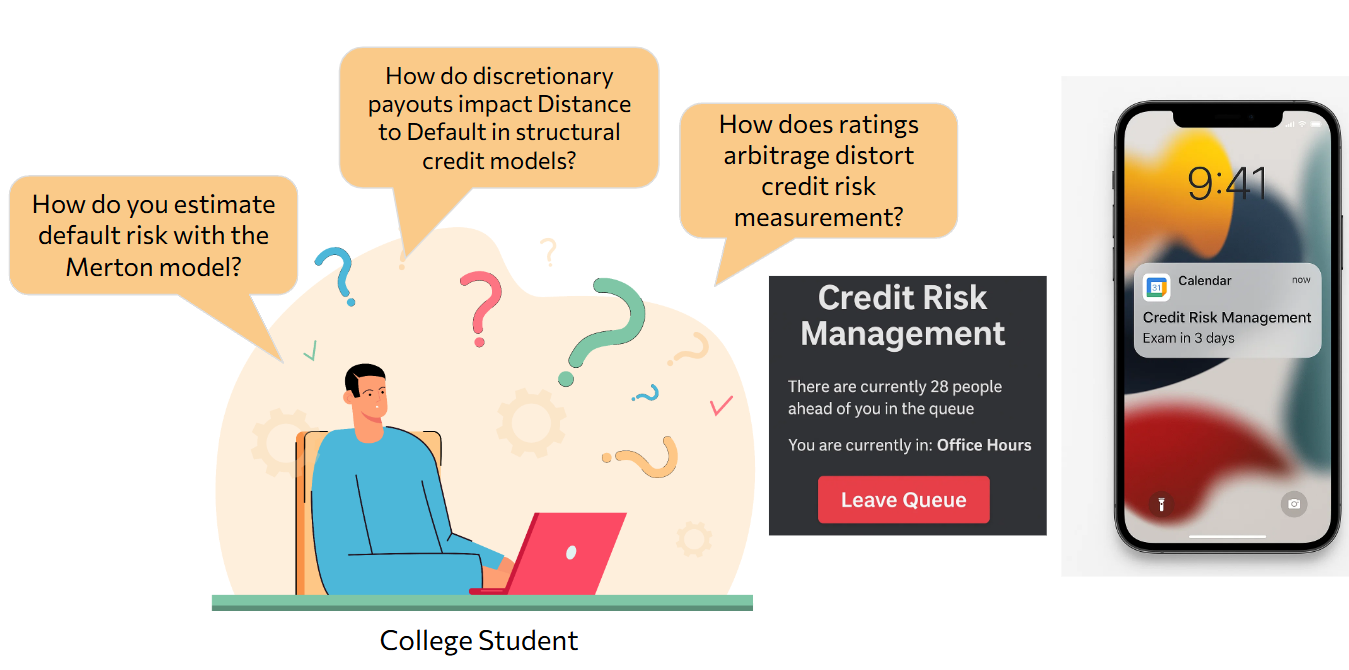}
    \caption{Financial education challenges addressed by AI tutors: affordability, scalability, and democratization of financial knowledge.}
    \label{fig:tutor_agent_challenges}
\end{figure}

Key features include \textbf{personalized learning} with 24/7 availability, tailored examples and practice questions, support for multiple educational levels, and step-by-step solution checking. The system provides \textbf{scalable online education} by handling massive student populations, offering real-time answers during lectures, and automated grading using advanced reasoning models.

Practical use cases demonstrate the versatility of financial AI tutors: students preparing for credit risk management exams receive immediate assistance without waiting time; software engineers quickly learn financial terms before meetings with FP\&A teams; online students pursuing independent research get guidance on unexplored areas and ongoing feedback; CFA candidates access expert-level explanations and tailored practice questions for complex topics.

\begin{figure}[t]
    \centering
    \includegraphics[width=\textwidth]{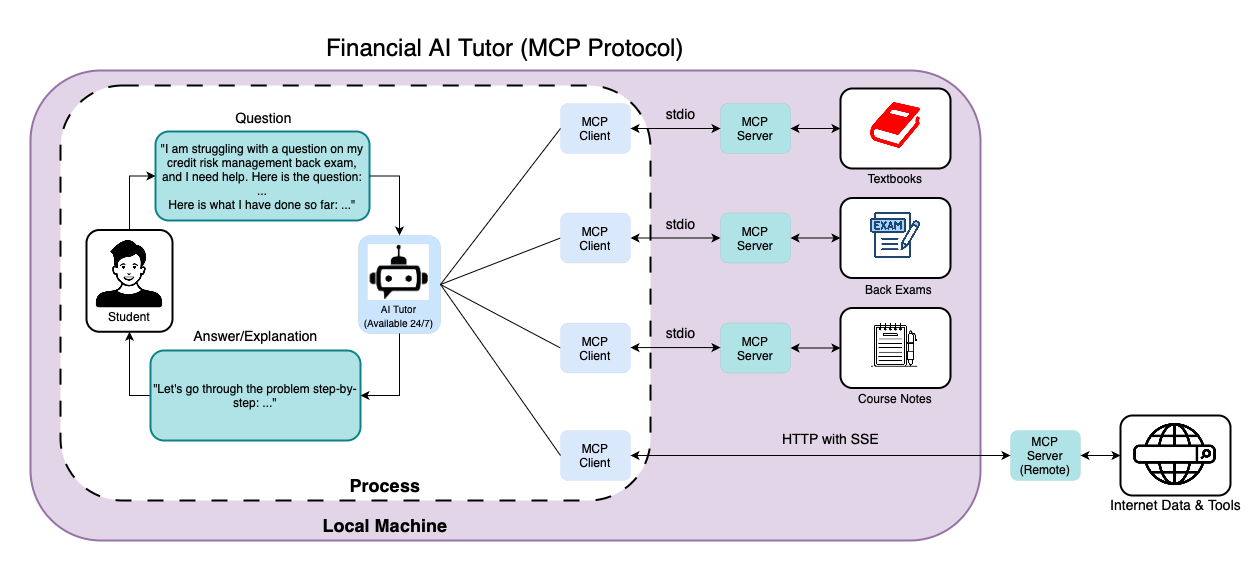}
    \caption{FinAI tutor agent using MCP protocol. The tutor provides instant answers to finance questions by accessing trusted resources like textbooks, past exams, and course notes.}
    \label{fig:tutor_agent_framework}
\end{figure}

Financial AI tutors can access trusted resources like textbooks, past exams, and course notes to provide instant answers to finance questions. They represent a significant advancement in democratizing financial education and making expert-level guidance accessible to a broader audience.

\subsection{FinSight Agent: A Metacognitive Multi-Agent System}

Moving beyond simple search and retrieval, the FinSight Agent (currently available in FINOS Labs) demonstrates how governance can be embedded directly into the agent's architecture and thinking process. Designed to analyze corporate earnings calls and integrating research from the ECC Analyzer research \cite{cao2024ecc}, FinSight employs a "Coordinator" agent that acts as a central brain, orchestrating specialized "Expert" agents for tasks such as Nuanced Sentiment Analysis, Significant Event Detection, and Volatility Prediction, leveraging tools and multi-modal data to complement its knowledge with multi-modal data. 

Crucially, FinSight was designed to employ Metacognitive Reasoning: the Coordinator self-assesses its confidence and risk levels as it generates responses. If a query is deemed high-risk or low-confidence, the system automatically triggers dynamic guardrails, including self-correcting loops and the ability to escalate the issue for human review. This approach illustrates a shift from passive guardrails to active, agentic self-governance, which influences our future evaluation and benchmarking approach.   

\begin{figure}
    \centering
    \includegraphics[width=\textwidth]{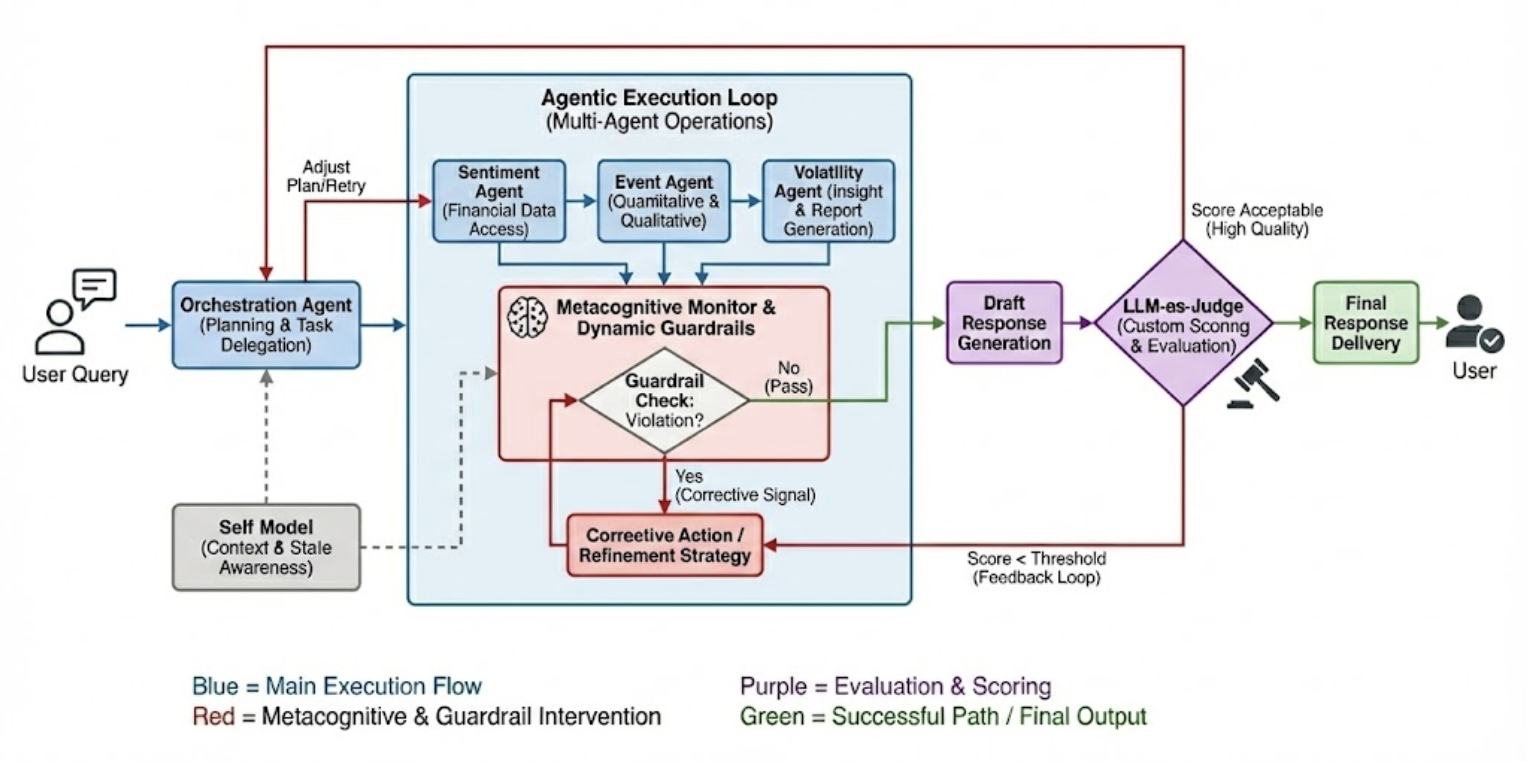}
    \caption{Cognitive architecture of the FinSight Agent, illustrating the Metacognitive execution flow. including task execution, evaluation and dynamic decisioning}
    \label{fig:placeholder}
\end{figure}

%Fig.~\ref{fig:use_case_overview} provides examples of how financial agents support analysis and decision-making processes.

\subsection{Social Media Monitor: A Case Study of GameStop Squeeze Event}

The GameStop (GME) short squeeze event in early 2021 highlighted the power of online communities in financial markets. It also generated a massive amount of discussion, questions, and misinformation. Prior work has analyzed the cascading outbreak mechanism of this event using network analysis and LLMs \cite{lin2024analyzing}. Financial LLMs can further be used to monitor community discussions, answer complex questions, and provide guidance.

\begin{figure}[t]
    \centering
    \includegraphics[width=\textwidth]{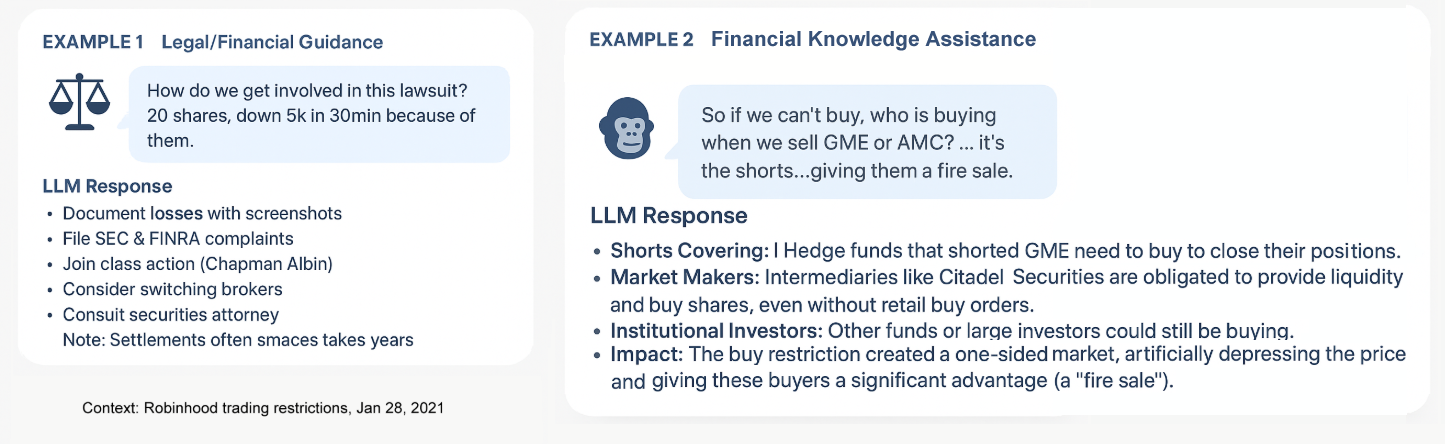}
    \caption{Use cases of FinLLMs: Analyzing community questions from the GameStop squeeze event. The model provides both legal/financial guidance and explains complex market mechanics.}
    \label{fig:gme_use_case}
\end{figure}

During the GameStop event, retail investors faced unprecedented market conditions and trading restrictions that generated widespread confusion and concern. Fig.~\ref{fig:gme_use_case} illustrates how FinLLMs can address questions that emerged from the community during this period.

\subsection{Sentiment Analyzer}

Bloomberg created BloombergGPT \cite{wu2023bloomberggpt} in 2023. It was a 50-billion parameter model trained on financial data. This was one of the first large language models made specifically for finance. It was trained on financial news, reports, and market data. The model learned to understand financial terms and context. The model could analyze \textbf{earnings call transcripts}, \textbf{financial news articles}, \textbf{regulatory filings}, and \textbf{financial jargon}.

BloombergGPT demonstrated that financial LLMs can perform sentiment analysis more effectively than general models and proved that domain-specific models are more effective for financial tasks. Financial LLMs could understand the complex language of finance.

Among BloombergGPT’s five public financial benchmarks, two tasks (FPB and FiQA SA) target sentiment analysis. In addition, five out of its twelve internal financial benchmarks are sentiment-related, representing nearly half of the total. These statistics illustrate BloombergGPT’s prioritization of sentiment analysis and reinforce the central role of sentiment understanding in financial LLMs.

\begin{figure}[t]
    \centering
    \includegraphics[width=\textwidth]{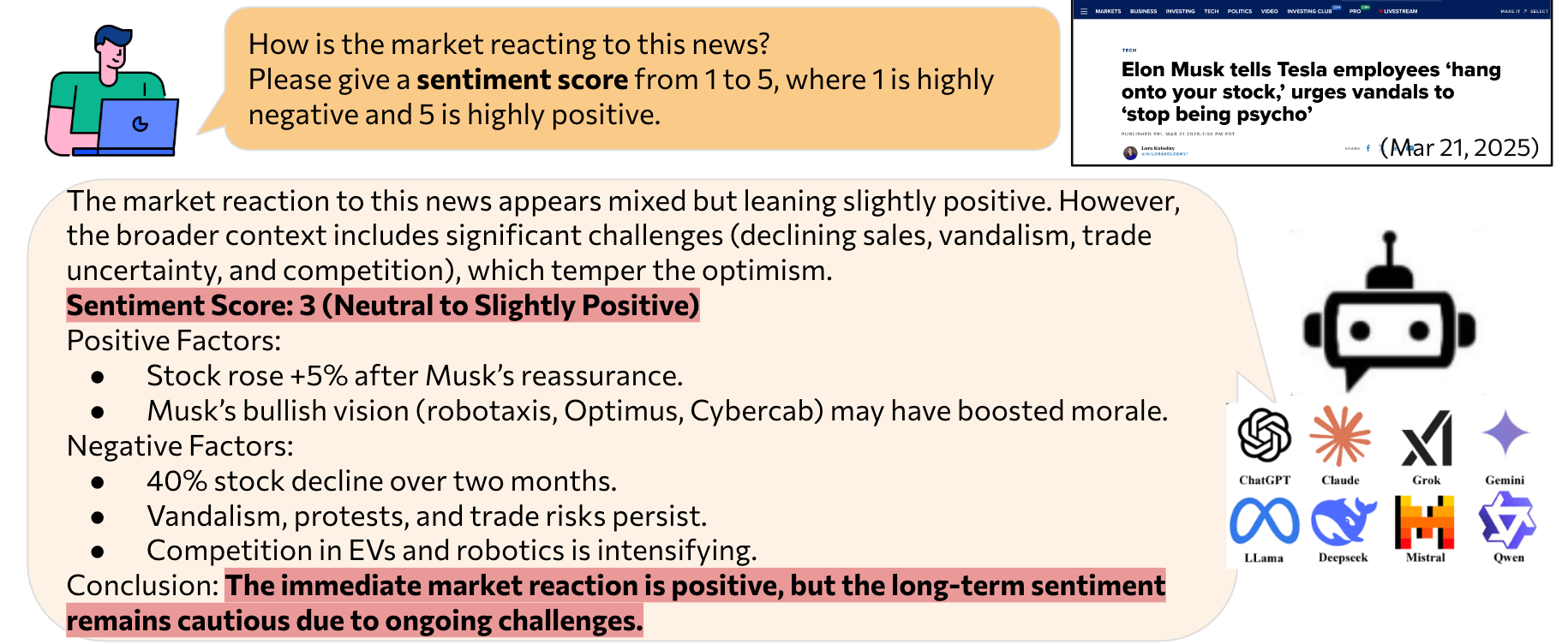}
    \caption{LLM-based sentiment analysis for generating alpha signals from financial news. The system analyzes market sentiment and provides trading signals with rationale.}
    \label{fig:sentiment_analysis}
\end{figure}

LLMs demonstrate significant potential for generating alpha signals from financial news through sentiment analysis. After retrieving financial information using search agents, users can extract trading signals for informed decision-making. For example, when analyzing CNBC news about Tesla, users can ask: "How is the market reacting to this news?" The LLM can interpret the article, analyze sentiment and tone, assign sentiment scores, and provide rationale. Fig.~\ref{fig:sentiment_analysis} shows how LLMs can conclude that while immediate market reaction appears positive, long-term sentiment remains cautious due to ongoing challenges. This demonstrates LLMs' capability in extracting actionable signals from unstructured financial documents for integration into trading strategies.

\subsection{SEC Analyzer}

SEC filings (e.g., 10-K, 10-Q) are the primary way for public companies to disclose information like revenue and cash flows to investors and regulators. SEC filings can be used by FinLLMs for investment and compliance purposes. However, to be effective on these documents, FinLLMs must have strong capabilities in long-context retrieval, numerical reasoning, table extraction and analysis, and temporal understanding of data.

Multiple datasets benchmarking these capabilities have been released. In 2021, TAT-QA \cite{zhu2021tat} introduced 16k+ questions from hybrid tabular and textual data, testing numerical reasoning and table analysis capabilities. After OpenAI released GPT-4 \cite{achiam2023gpt}, FinanceBench \cite{islam2023financebench} introduced 10k+ QA pairs from SEC filings, which focus on long-context retrieval, numerical reasoning, and table analysis. GPT-4-Turbo showed mixed performance across different retrieval and context settings, motivating research into better long-context and retrieval-augmented approaches. Recently, SECQUE \cite{yoash2025secque} introduced 565 expert-written questions, spanning four key categories: comparison and trend analysis, ratio analysis, risk factors, and analyst insights. SECQUE also contributed SECQUE-judge, an LLM-as-a-judge with strong human alignment for scoring.

\begin{figure*}[t]
    \centering
    \includegraphics[width=\textwidth]{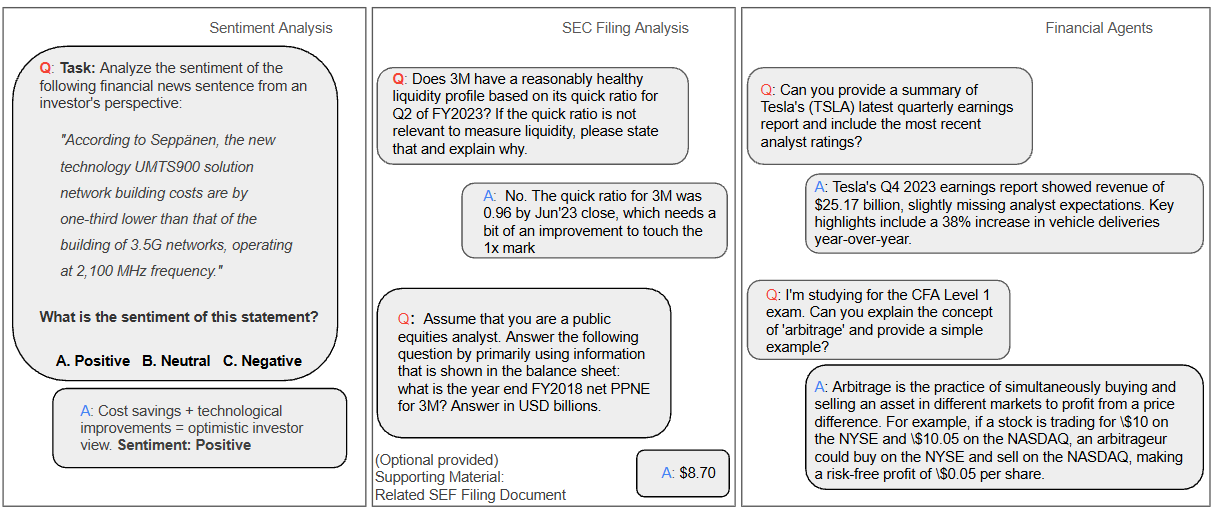}
    \caption{Overview of financial LLM use cases: Sentiment Analysis, SEC Filing Analysis, and Financial Agents, each illustrated with example questions and model answers.}
    \label{fig:use_case_overview}
\end{figure*}

\subsection{ECC Analyzer}

The earnings conference call (ECC) is a teleconference or webcast held quarterly by a public company. Stakeholders (including analysts, investors, and the media) participate to obtain the company's latest financial status. First, the company's CEO/CFO highlights the quarterly financial status, strategic initiatives, and forward-looking plans. Then, analysts and investors ask questions in the Q\&A sessions. The release of ECCs is correlated with market reactions, making them an important resource for analyzing market changes.

\begin{figure}[htbp]
    \centering
    \includegraphics[width=0.75\textwidth]{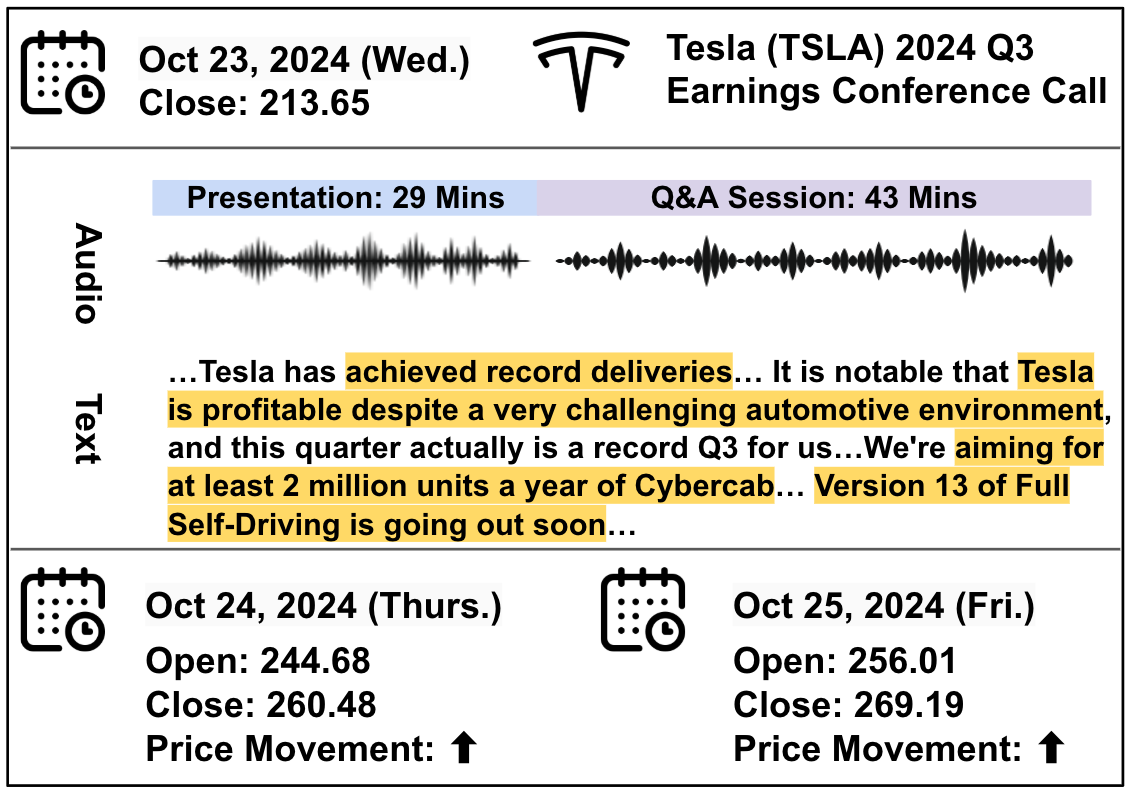}
    \caption{Tesla 2024 Q3 earnings conference call example showing 72-minute duration with 29-minute CEO presentation followed by 43-minute Q\&A session.}
    \label{fig:ecc_example}
\end{figure}

An example of Tesla 2024 Q3 ECC, as shown in Fig.~\ref{fig:ecc_example}, has 72 minutes. The call includes 29 minutes presentation by Tesla CEO Elon Musk, followed by 43 minutes Q\&A session. First, CEO Elon Musk summarized Tesla's Q3 revenue and car production status and underscored Tesla's ongoing strategy to hasten the global transition to sustainable energy. In the end, Musk reiterated Tesla's preparations for introducing more affordable models. In the Q\&A session, Tesla's executive team responded to questions about product research and development, upcoming product plans, Tesla's Full Self-Driving offerings, etc. Owing to good revenue performance and car production, Tesla's stock price sustained an upward trend in the following two days. The entire ECC is saved as a ".mp3/.wav" audio file, and the corresponding transcript is also recorded. Both audio and text data can be accessed or analyzed by the public.

ECC data presents unique challenges: Long Audio - The audio of the ECC is long, typically lasting 30-60 minutes, which makes it challenging to analyze the entire audio in one go. Long Transcript - The transcript of the ECC is also long, often exceeding 200+ sentences, which requires efficient processing and analysis methods.

\begin{figure}[htbp]
    \centering
    \includegraphics[width=\textwidth]{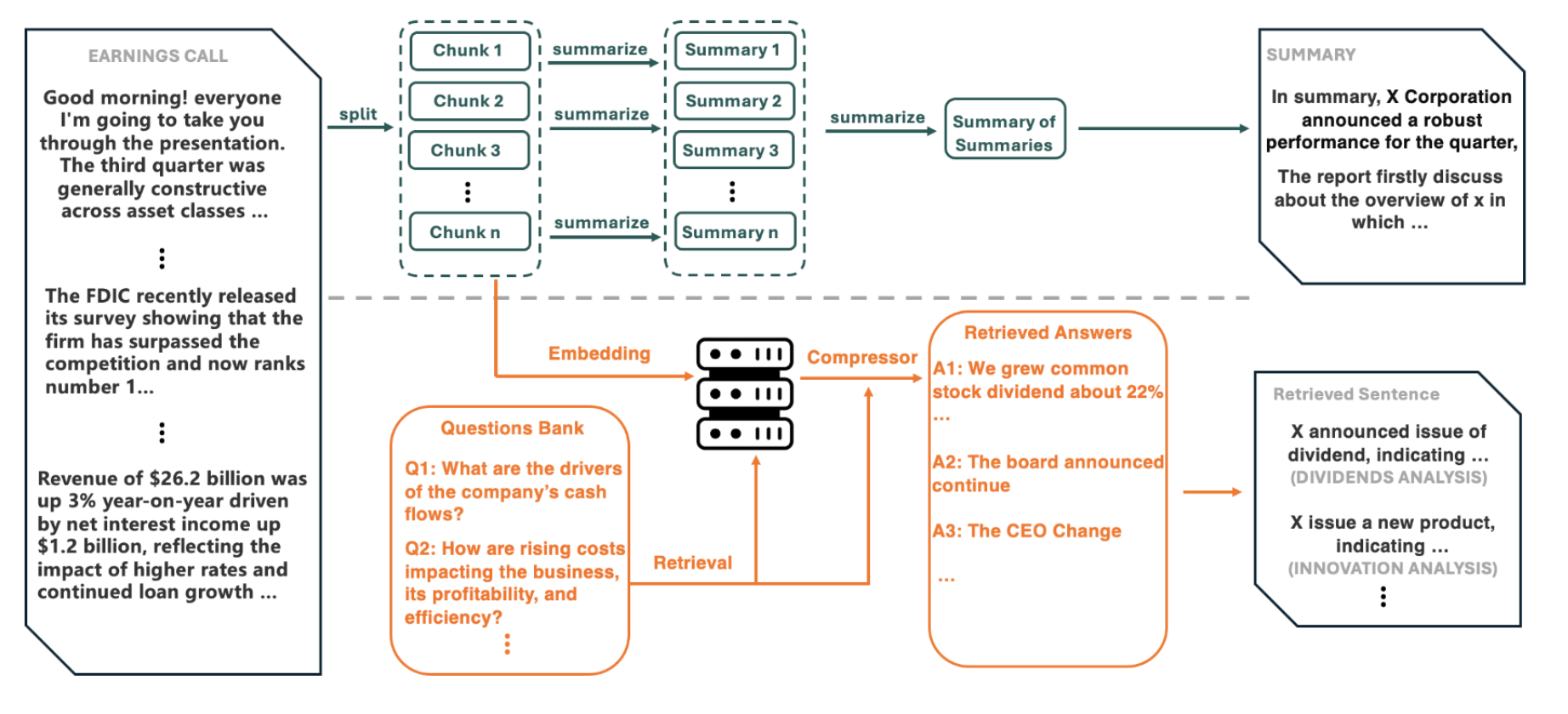}
    \caption{ECC Analyzer framework \cite{cao2024ecc} for extracting trading signals from earnings conference calls using large language models for stock volatility prediction.}
    \label{fig:ecc_analyzer}
\end{figure}

The ECC Analyzer, shown in Fig.~\ref{fig:ecc_analyzer}, addresses these challenges by extracting trading signals from earnings conference calls using large language models for stock volatility prediction \cite{cao2024ecc}.

\section{FinAI Governance: Building Guardrails for Real-World Deployment  }
\label{sec:benchmarking}

\subsection{Governance Requirements}
\label{subsubsec:finos_framework}

Generative AI is reshaping financial services by enhancing products, client interactions, and productivity. However, challenges like \textbf{hallucinations} and \textbf{model unpredictability} make safe deployment complex. The \href{https://air-governance-framework.finos.org/}{Linux Fundation AI Governance Framework} provides a comprehensive collection of risks and mitigations for Generative AI solutions in financial services. These risks can be further identified in the application of FinAgents in financial scenarios, such as private financial data leakage and LLM hallucinations when preparing SEC filings. Retrieval-augmented generation can help reduce hallucinations and ensure financial and regulatory information is up-to-date. Zero-knowledge proof techniques \cite{sun2024zkllm} can help protect data privacy and the intellectual properties of LLMs.

Furthermore, financial institutions increasingly need to consider compliance with new regulatory frameworks around the use of Generative AI models, such as the EU AI Act which requires detailed documentation of model training and data provenance. In aspects related to model governance, we propose to leverage the Model Openness Framework \cite{mof2024white} which provides a standardized schema for this compliance with financial agents prioritizing MOF Class I models to ensure:
\begin{itemize}[leftmargin=*]
    \item Data Lineage: Verifying that no Material Nonpublic Information (MNPI) or biased datasets were used during pretraining.
    \item Reproducibility: Ensuring that a model's financial reasoning can be reproduced and audited by internal risk teams.
    \item Legal Certainty: clarifying licensing rights for commercial financial applications, for example by using the \href{https://openmdw.ai/}{OpenMDW} license.
\end{itemize}

\subsection{Operational Evaluation and Governance: The AgentOps Framework}

While static benchmarks measure model capability, increasingly we see real-world governance that requires evaluating agents as complex, autonomous systems. We propose adopting an AgentOps framework \cite{dashet2025fromlabtolife} that splits evaluation into two distinct loops to ensure reliability in "the wild", based on the methodology defined by Jabbour et al. \cite{jabbour2025evaluationframeworkaisystems}: it spans the AI system from development and testing (the inner loop) to production (observability, feedback, and governance (in the outer loop) 
\begin{itemize}[leftmargin=*]
    \item The Inner Loop (Development and Testing): This phase focuses on "glass box" evaluation, where the agent's reasoning trajectory is transparent. Unlike traditional "black box" testing which only scores the final answer, inner loop evaluation utilizes Trajectory Tracing to analyze the decision path, tool usage, and self-correction steps. Key metrics include Tool Selection Accuracy, Context Adherence, and Step-wise Latency.
    \item The Outer Loop (Production and Observability): Once deployed, continuous evaluation is maintained through audit trails and feedback loops. This typically involves a range of dynamic evaluations including a "LLM-as-a-Judge," where a high-capacity model (potentially with human-in-the-loop involved in reviewing its accuracy over time) scores live interactions for safety, ethics, and logic violations. This ensures that unwanted "non-deterministic" behaviors—common in agentic workflows—are detected and corrected in real-time.
    \item The Feedback Loop: One of the most important aspects in this approach is closing the gap between Online and Offline evaluations. The approach to reconcile evaluations across both loops is to take those real-world failures and edge cases identified in Online Evals stage and use the relevant observability data such as agent traces to contribute both real and synthetic data back to the Offline Evals process. This ensures that real-world problems become part of our automated regression suite and making our offline evaluation a more accurate predictor of online performance.
\end{itemize}

\begin{figure}
    \centering
    \includegraphics[width=\textwidth]{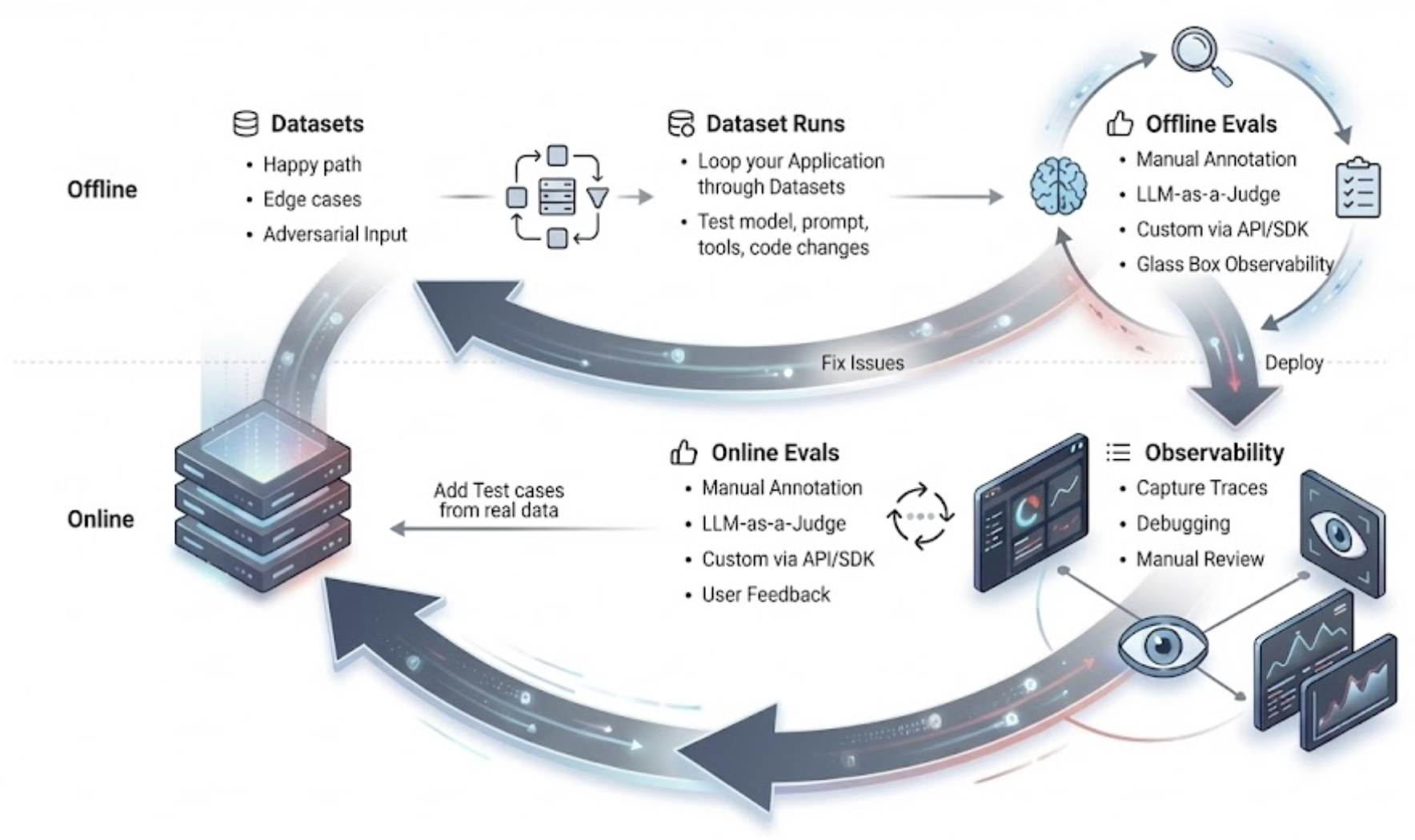}
    \caption{Our AgentOps framework: Unlike traditional software that follows a deterministic path through code execution, agents loop and iterate with non-deterministic thinking. Our AgentOps strategy covers the entire evaluation lifecycle: the 'inner loop' during development, where we visualize the agent's decision paths (trajectory analysis) through evaluation of traces, and test specific outcomes and guardrails; and the 'outer loop' in production including custom performance monitoring in real-time}
    \label{fig:placeholder}
\end{figure}

% \section{Collaborative Projects at Linux Foundation}
% \label{sec:collaboration}

% We collaborative projects at Linux Foundation and focus on developing frameworks for responsible AI deployment in finance. This includes the Model Openness Framework (MOF) and governance considerations for financial LLMs.

% \subsection{Model Openness Framework (MOF)}
% \label{subsec:mof}

% The Generative AI Commons at the LF AI \& Data Foundation developed the Model Openness Framework (MOF). This framework evaluates and classifies the \textbf{completeness} and \textbf{openness} of machine learning models. It assesses which components of the model development lifecycle are publicly released and under what licenses.

% The framework addresses three main categories of risks. \textbf{Operational risks} include hallucination and inaccurate outputs, foundation model versioning, non-deterministic behavior, and bias and discrimination. \textbf{Security risks} cover information leakage to vector stores, data poisoning, and prompt injection. \textbf{Regulatory and compliance risks} include information leakage to hosted models, regulatory compliance oversight, and intellectual property concerns.

% The framework provides two types of mitigations. \textbf{Preventative controls} include data filtering from external knowledge bases, user/app/model firewalling, system acceptance testing, and AI model version pinning. \textbf{Detective controls} include AI data leakage prevention and detection, AI system observability, human feedback loops, and providing citations for AI-generated information.

\subsection{Model Openness Framework}
\label{sec:MOF}

\begin{figure}[t]
    \centering
    \includegraphics[width=\textwidth]{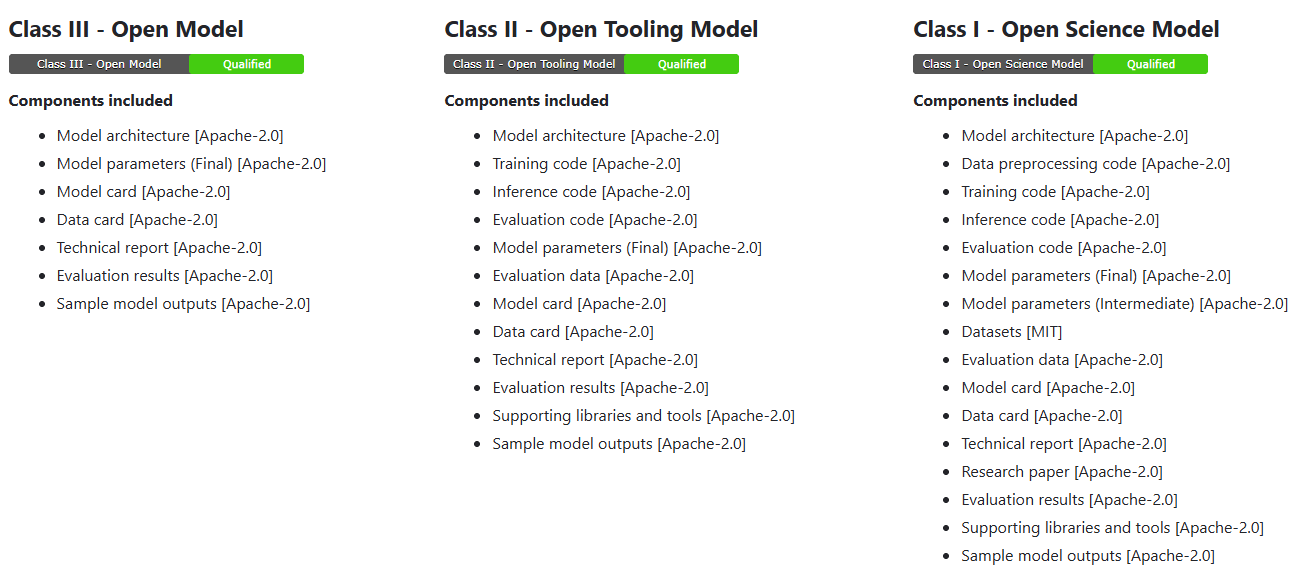}
    \caption{Model Openness Framework (MOF) classification system: Open Model Class 1, Class 2, and Class 3, based on the completeness and accessibility of model components.}
    \label{fig:mof_classes}
\end{figure}

The Generative AI Commons at the LF AI \& Data Foundation has developed the \textbf{Model Openness Framework (MOF)} \cite{mof2024white}, which evaluates and classifies the completeness and openness of LLMs. With the rise of AI democratization, more and more models are claimed to be open. However, model users often face uncertainty about which specific components are truly open and do not understand the associated licenses. As a result, "openwashing" behavior becomes common among AI models. To address this problem, the MOF identifies 17 components along the lifecycle of model development, including code, data, and documentation, each with suggested open licenses. It classifies the completeness and openness of models into three levels: Class III Open Model, Class II Open Tooling, and Class I Open Science, as shown in Fig.~\ref{fig:mof_classes}. With the MOF, model users can better understand what model producers provide, what model components are open, and how to use and distribute models under open licenses. This will enhance the healthy and standardized development of FinAgents built on open models. 

\section{Conclusion}
\label{sec:conclusion}

This paper presents a comprehensive framework for understanding the evolving lifecycle of financial LLMs through three critical stages: Exploration (2023), Readiness (2024), and Governance (2025). We introduce the Open FinLLM Leaderboard as a standardized benchmarking suite that enables systematic evaluation and comparison of financial AI systems. This work contributes to the development of more robust and reliable financial AI systems. Future work will focus on expanding governance frameworks and enhancing evaluation methodologies to address emerging challenges in building a reliable FinAI ecosystem.

%\section*{Acknowledgment}

% Bibliography
\bibliographystyle{IEEEtran}
\bibliography{references}

% Appendices
\appendix
\clearpage

\end{document}